%% file: conference_101719.tex
\documentclass[a4paper, 10pt, conference]{IEEEtran}
\IEEEoverridecommandlockouts
\input{macros.tex}

\usepackage{cite}
\usepackage{amsmath,amssymb,amsfonts}
\usepackage{lipsum}
\usepackage{graphicx}
\usepackage{subcaption}
\usepackage[a4paper]{hyperref}
\usepackage{textcomp}
\usepackage{xcolor}
\def\BibTeX{{\rm B\kern-.05em{\sc i\kern-.025em b}\kern-.08em T\kern-.1667em\lower.7ex\hbox{E}\kern-.125emX}}

\usepackage{tikz}
\usepackage{pgfplots}
\pgfplotsset{compat=newest}
\pgfplotsset{plot coordinates/math parser=false}
\newlength\figH  %figureheight
\newlength\figW  %figurewidth
\usetikzlibrary{quotes,arrows.meta}
\usepackage{physics}
\usepackage{amssymb}
\usepackage{amsmath}
\usepackage{amsthm}
\usepackage{algpseudocode}
\usepackage{algorithm,caption}
\usepackage[english]{babel}

\usepackage{graphicx}
\usepackage{wrapfig}
\usepackage{caption}
\usepackage{array}
\usepackage{multirow}
\usepackage{arydshln}
\usepackage{color}
\usepackage{xcolor}
\usepackage{tcolorbox}
\usepackage[framemethod=tikz]{mdframed}
\newcommand{\tikznode}[2]{%
	\ifmmode%
	\tikz[remember picture,baseline=(#1.base),inner sep=0pt] \node (#1) {$#2$};
	\else
	\tikz[remember picture,baseline=(#1.base),inner sep=0pt] \node (#1) {#2};%
	\fi}

\definecolor{darkgreen}{rgb}{0,.39,0}

\definecolor{darkred}{rgb}{0.4,0,0}

\definecolor{dnared}{rgb}{1,.27,0} 
\definecolor{dnablue}{rgb}{0,1,1}
\definecolor{dnapurple}{rgb}{0.82,0.13,0.56}
\definecolor{dnagreen}{rgb}{0,1,0}

\usepackage{tikz}

\usetikzlibrary{%
	arrows,%
	automata,%
	backgrounds,%
	calc,%
	patterns,%
	plotmarks,%
	positioning,%
	shapes.geometric,%
	shapes,%
	snakes,%
	decorations.pathmorphing,%
	decorations.shapes,%
	graphs,%
	chains,%
	arrows.meta, %
}

\usepackage{verbatim}
\usepackage{tabularx}
\usepackage{lipsum}
\usepackage{stackengine}
\usepackage{nicefrac}

\usepackage{svg}
\definecolor{bostonuniversityred}{rgb}{0.8, 0.0, 0.0}

\usepackage{xcolor}

\usepackage{mathdots}

\usepackage{afterpage}
\usepackage{lipsum}
\usepackage{fancyhdr}

\fancypagestyle{IEEEtitlepagestyle}{
    \fancyhf{} % clear all header and footer fields
    \fancyfoot[L]{\vspace{12pt}%
    \footnotesize
    \copyright This work has been submitted to the IEEE for possible publication. Copyright may be transferred without notice, after which this version may no longer be accessible.}
}

\begin{document}

\title{Bilinear Precoder Based Efficient Rate Splitting Method in FDD Systems}

\author{\IEEEauthorblockN{Sadaf Syed*, Donia Ben Amor*, Michael Joham, Wolfgang Utschick }
\IEEEauthorblockA{{School of Computation, Information and Technology, Technical University of Munich, Germany}\\
Emails: \{sadaf.syed, donia.ben-amor, joham, utschick\}@tum.de}
\thanks{* These authors contributed equally to this work.
}
}

\maketitle

\begin{abstract}
In this work, we propose a low-cost rate splitting~(RS) technique for a multi-user multiple-input single-output~(MISO) system operating in frequency division duplex~(FDD) mode. The proposed iterative optimisation algorithm only depends on the second-order statistical channel knowledge and the pilot training matrix. Additionally, it offers a closed-form solution in each update step. This reduces the design complexity of the system drastically as we only need to optimise the precoding filters in every coherence interval of the covariance matrices, instead of doing that in every channel state information (CSI) coherence interval. Moreover, since the algorithm is based on closed-form solutions, there is no need for interior point solvers like CVX, which are typically required in most state-of-the-art techniques.  
\end{abstract}

\section{Introduction}
Massive multiple-input-multiple-output~(MaMIMO) systems are foreseen to be adopted in the future wireless communication systems due to their high spectral and energy efficiencies. However, acquisition of the perfect channel state information~(CSI) is very difficult due to the small channel coherence interval. The imperfect CSI can degrade the system's performance, especially in the high transmit power regime. Rate splitting~(RS) has received a considerable attention for B5G/6G systems recently and its benefits are seen in scenarios with both perfect and imperfect CSI~\cite{rs1, rs3, rs4, rs5, amor2024efficient, bruno, amor2023rate}. RS mitigates the interference effectively by splitting the messages of the users into common and private parts. The common parts are jointly encoded into a common stream, which is decoded by all users, and the private parts are encoded into the respective private streams, which are decoded only by the corresponding users. Therefore, RS allows the users to partially decode the interference and partially treat it as noise. Each user decodes the common message firstly and then applies
successive interference cancellation to remove the interference due to the common part. 
\par In this work, we address the challenging scenario of a frequency-division-duplex~(FDD) multi-user (MU) multiple-input-single-output~(MISO) system, where the base station~(BS) only has access to imperfect CSI. Unlike the time-division-duplex (TDD) systems, where channel reciprocity is assumed between the uplink (UL) and downlink (DL) channels, FDD systems lack this property. Therefore, the BS must explicitly estimate the DL channels by sending pilots to the users and relying on their feedback to obtain CSI. Due to the large number of antennas at the BS, especially in a MaMIMO system, the number of DL training pilots used for channel estimation is generally considerably less than that of BS antennas, resulting in incomplete CSI. Most of the existing algorithms in the literature employing RS strategy require optimisation of the common and private beamforming vectors in every CSI coherence interval \cite{bruno, amor2024efficient, rs1, rs3, rs4, rs5}. However, this is computationally very expensive because of the limited channel coherence interval. In this work, we propose a low-complexity algorithm which exploits the statistical knowledge of the channels to reduce the system design complexity. Similar to the algorithm in \cite{amor2023rate}, the proposed method employs the lower bound on the achievable sum-rate of the users as the figure of merit, which is based on the worst-case noise bound~\cite{medard}. We consider a setup where the covariance matrices of the channels are known perfectly, however, the accurate knowledge of the instantaneous CSI is not available at the BS. In order to reduce the hardware complexity at the users' side, one-layer RS is implemented, i.e., only one common stream is transmitted to all users. In order to formulate an optimisation problem which only depends on the second-order channel statistics, the bilinear precoding~(BP) approach of~\cite{amor2020bilinear} is employed, similar to \cite{amor2023rate}. This precoder consists of a linear transformation of the received training signal via a deterministic transformation matrix, which only depends on the covariance matrices of the channels and the pilot training matrix. Since the structure of the channels varies slowly, the second-order channel statistics remain constant for many CSI coherence intervals. Hence, it is relatively easier to obtain accurate information of the channel covariance matrices through long-term observation. The transformation matrices designed on the basis of the second-order channel statistics do not need to be updated regularly, thereby reducing the complexity of the algorithm. The algorithm in \cite{amor2023rate} also uses a BP based RS strategy, where the optimisation problem is solved independently for the common and the private beamforming vectors in each iteration. The common rate optimisation in \cite{amor2023rate} uses the SINR increasing algorithm of~\cite{hunger2007design}. However, the optimisation of the private and the common precoders is coupled, which makes their independent optimisation suboptimal. Additionally, in order to compute the optimal split of the transmit power between the common and the private streams, an additional power allocation optimisation step needs to be performed in \cite{amor2023rate}, which further increases the computational complexity of the algorithm.
\par Our contributions in this work are summarised as follows: We employ a BP based RS strategy where the optimisation process is performed only once in every covariance matrix coherence interval. In the absence of complete CSI, a closed-form expression of the actual sum-rate of the users cannot be computed. Hence, we employ the lower bound of the achievable sum-rate \cite{medard}. We develop a heuristic algorithm similar to \cite{amor2024efficient}, where the common and the private precoders are optimised jointly. Moreover, all the update steps in each iteration have a closed-form solution, which circumvents the need of using interior-point solvers like CVX \cite{grant2014cvx}. Simulation results validate the efficiency of our algorithm in terms of the computational time and its effectiveness compared to the non-RS setup in the presence of imperfect CSI.   
\section{System Model and Problem Formulation}
This paper investigates the downlink (DL) of a multi-user MISO frequency division duplex~(FDD) communication system with one base station~(BS) having $M$ antennas and serving $K$ single-antenna users. The channel between the BS and the $k$-th user is denoted by $\bfh_{k}\in \mathbb{C}^{M}$, and it is assumed to be circularly symmetric, complex Gaussian distributed with $\bfh_{k}\sim\nc({\bf{0}}, \bfC_{k})$, where $\bfC_k = \Ex[\bfh_k \bfh_k^{\Hm}]$ is the channel covariance matrix. The BS sends $T_{\text{dl}}$ orthogonal pilot sequences to all the users for channel estimation during the training phase. The $T_{\text{dl}}$ training symbols at the $k$-th user constitute the observation vector $\bfy_k$, which is given by  
\begin{align}
    \bfy_k = \Phimat^{\Hm}\bfh_k + \bfn_k
\end{align}
where $\Phimat \in \mathbb{C}^{M\times T_{\text{dl}}}$ denotes the pilot matrix with $T_{\text{dl}}$ orthogonal columns, and $\bfn_k \sim\nc({\bf{0}}, \sigma_k^2\:{\mathbf{I}}_{T_{\text{dl}}})$ is the DL training noise. In order to reduce the training overhead, especially in the case of massive MIMO systems, it is assumed that $T_{\text{dl}} \ll M$. Thus, the BS only has access to incomplete CSI due to the resulting systematic error. 
\par In the data transmission phase, a single-layer RS approach is applied at the BS such that the message intended for a user is split into one common message and one private message. The common messages of all the users are combined and encoded into a super-common message $s_\text{c}$, whereas the private messages are independently encoded into $s_1, \cdots, s_K$. We denote $\bfs = [s_\text{c}, s_1, \cdots , s_K]^\Tm \in \mathbb{C}^{K+1}$ as the data stream vector, $\bfp_\text{c} \in \mathbb{C}^{M}$ as the common precoding vector and $\bfp_1, \cdots, \bfp_K \in \mathbb{C}^{M}$ as the private beamforming vectors. Each stream in $\bfs$ is assumed to be zero-mean with unit variance, i.e., $\Ex[{\bfs \bfs^{\Hm}}]=\mathbf{I}$. The total received signal at the $k$-th user reads as
\begin{align}
    r_{k} =\bfh_{k}^{\Hm}\bfp_{\text{c}}\:s_{\text{c}} + \sum\limits_{i = i }^K \bfh_{k}^{\Hm}\bfp_{i}\:s_{i} + z_{k}
\end{align}
where $z_{k} \sim\mathcal{N_\mathbb{C}}(0, 1) $ denotes the additive white Gaussian noise (AWGN) at the user $k$.
 \par The beamforming vectors at the BS are designed such that they only depend on the second-order channel statistics and the noisy observations $\bfy_k$. To this end, we employ the bilinear precoder (BP) \cite{amor2020bilinear} based RS approach, as done in \cite{amor2023rate}. According to \cite{amor2023rate}, the common and private precoders can be written as  
 \begin{align}
   \bfp_{\ct} &= \bfA_{\ct} \bfy \label{eqn3} \\ 
   \bfp_k &= \bfA_k \bfy_k \label{eqn4}
\end{align}
where $\bfy = [\bfy_1^{\Tm}, \cdots, \bfy_K^{\Tm}]^{\Tm}$. The matrices $\bfA_{\ct} \in \mathbb{C}^{M\times K\: T_{\text{dl}}}$ and $\bfA_k \in \mathbb{C}^{M\times  T_{\text{dl}}}$ are deterministic transformation matrices which only depend on the second-order channel statistics and the pilot matrix $\Phimat$. Note that these matrices need to be designed only once in every coherence interval of the covariance matrices, i.e., we do not need to optimise the beamforming vectors in every channel coherence interval, hence, reducing the system's computational complexity. The common precoder in \eqref{eqn3} is given by  $ \bfp_{\ct} =  \sum\nolimits_{k=1}^{K}\bfA_{\ct,k}\bfy_k$ with $\bfA_{\ct} = [\bfA_{\ct,1}, \cdots, \bfA_{\ct,K}]$. This choice is motivated by the multicast transmission as done in \cite{amor2023rate}. 
\par Because of the imperfect CSI, we cannot compute a closed-form expression of the achievable sum-rate of the users. Instead of that, we employ a lower bound on the users' sum-rate based on the worst-case error, which is extensively used in the massive MIMO literature~\cite{medard}. The lower bounds for the common and private rates of the $k$-th user are given by $R_{\ct,k}^{{\mathrm{lb}}} = \log_{2}(1 + \gamma_{\ct,k}^{\mathrm{lb}})$ and $R_{\text{p},k}^{{\mathrm{lb}}} = \log_{2}(1 + \gamma_{\text{p},k}^{\mathrm{lb}})$, where $\gamma_{\ct,k}^{\mathrm{lb}}$ and $\gamma_{\text{p},k}^{\mathrm{lb}}$ can be expressed as
\begin{align}
\label{lb_c}
\gamma_{\ct,k}^{\mathrm{lb}} &= \dfrac{|\mathop{{}\mathbb{E}}[\bfh_k^{\Hm}\bfp_{\ct}]|^2}{\mathrm{var}({\bfh_k^{\Hm}\bfp_{\ct}})  + \sum\nolimits_{\substack{j = 1 }}^{K}{\mathop{{}\mathbb{E}}[|{{\bfh_k^{\Hm}}{\bfp_j}}|^2] + 1}}
\end{align}
\begin{align}
\label{lb_p}
\gamma_{\text{p},k}^{\mathrm{lb}} &= \dfrac{|\mathop{{}\mathbb{E}}[\bfh_k^{\Hm}\bfp_k]|^2}{\mathrm{var}({\bfh_k^{\Hm}\bfp_k})  + \sum\nolimits_{\substack{j = 1 \\ j \neq k}}^{K}{\mathop{{}\mathbb{E}}[|{{\bfh_k^{\Hm}}{\bfp_j}}|^2] + 1}}
\end{align}
where $\mathrm{var}(.)$ denotes the variance operator.
Evaluating the terms in \eqref{lb_c} and \eqref{lb_p} as done in \cite{amor2023rate}, we get
\begin{align}
\label{eqn7}
\gamma_{\ct,k}^{\mathrm{lb}} & = \dfrac{\left|{\bfz_k^{\Hm}\bfa_{\ct}}\right|^2}{ \bfa_{\ct}^{\Hm}\bfZ_k\bfa_{\ct} 
 + \left|{\bfq_k^{\Hm}\bfa_k}\right|^2 + \sum\nolimits_{j=1}^{K}{\bfa_j^{\Hm}\bfQ_{j,k} \bfa_j} + 1} \\
\gamma_{\text{p},k}^{\mathrm{lb}} & =  \dfrac{\left|{\bfq_k^{\Hm}\bfa_k}\right|^2}{ \sum\nolimits_{j=1}^{K}{\bfa_j^{\Hm}\bfQ_{j,k} \bfa_j} + 1} \label{eqn8}
\end{align}
where $\bfa_{\ct} = \text{vec}(\bfA_{\ct})$, $\bfa_k = \text{vec}(\bfA_k$), and $\bfz_k$, $\bfq_k$, $\bfZ_k$ and $\bfQ_{j,k}$ depend on the second-order channel statistics and $\Phimat$, as derived in the Appendix. Hence, the sum-rate maximisation problem under the transmit power constraint is given by 
    \begin{align}
    &\!\max\limits_{\bfA_{\ct}, \bfA_1, \cdots, \bfA_K }  &  & \sum\limits_{i=1}^{K}\log_2\left(1 + \gamma_{\text{p},i}^{\mathrm{lb}}  \right) \: +  \:\!\min\limits_{k} \log_2\left(1 + \gamma_{\ct,k}^{\mathrm{lb}}\right) \nonumber \\
    & \quad \text{s.t.}  &     &  \sum\limits_{i=1}^{K}\mathop{{}\mathbb{E}}\left[\|{\bfp_i}\|^2 \right] + \mathop{{}\mathbb{E}}\left[\|{\bfp_{\ct}}\|^2 \right] \leq P_{\text{t}}.   \tag{P1} \label{P1}
    \end{align}
The inner minimisation over the user indices $k$ in \eqref{P1} ensures that the common message can be decoded by all users. Using the expressions of $\bfp_{\ct}$ and $\bfp_k$ from \eqref{eqn3} and \eqref{eqn4}, the DL power constraint at the BS can be written in closed-form as
\begin{align}
    \sum\limits_{i=1}^{K} \bfa_i^{\Hm}\bfF_i \bfa_i + \bfa_{\ct}^{\Hm}\bfF \bfa_{\ct} \leq P_{\text{t}} \label{eqn9}
\end{align}
where $\bfF_i = \bfC_{\bfy_i}^{\Tm} \otimes {\mathbf{I}}_M$, $\bfF = \text{blkdiag}(\bfF_1, \cdots, \bfF_K)$, $\otimes$~denotes the Kronecker product, and $ \bfC_{\bfy_i} = \Ex[\bfy_i \bfy_i^{\Hm}] = \Phimat^{\Hm} \bfC_i \Phimat \:+ \: \sigma_i^2 \:{\mathbf{I}}_{T_{\text{dl}}} $. 
\section{Simplification of the Optimisation Problem}
It is difficult to solve \eqref{P1} in closed-form, especially due to the search of the minimum common rate among all users. The algorithm in \cite{amor2023rate} solves \eqref{P1} by optimising the common and private precoders independently in each iteration. It is evident from \eqref{eqn7} that the common rate also depends on the private precoding transformation matrices $\bfA_k$. Hence, the optimisations of $\bfA_{\ct}$ and $\bfA_k$ are coupled. In this work, we employ a low complexity heuristic approach which jointly optimises the common and private beamforming vectors. This heuristic algorithm is motivated by the RS approach in \cite{amor2024efficient}, where the beamforming vectors are optimised in every channel coherence interval. In this work, we modify the algorithm in \cite{amor2024efficient} for the case of BP using the second-order channel statistics. The algorithm first solves \eqref{P1} for every choice of user index $k_{\ct} \in \{1, \cdots, K \}$, which tackles the inner minimisation block. Therefore, the corresponding optimisation problem reads as
\begin{align}
    &\!\max\limits_{\bfa_{\ct}, \bfa_1, \cdots, \bfa_K }  &  & \sum\limits_{i=1}^{K}\log_2\left(1 + \gamma_{\text{p},i}^{\mathrm{lb}}  \right) \: +  \:\log_2\left(1 + \gamma_{\ct,\kc}^{\mathrm{lb}}\right)  \nonumber \\
     & \quad \text{s.t.}  &     &   \sum\limits_{i=1}^{K} \bfa_i^{\Hm}\bfF_i \bfa_i + \bfa_{\ct}^{\Hm}\bfF \bfa_{\ct} \leq P_{\text{t}}. \tag{P2} \label{eqn10}
    \end{align}
This results in $K$ different choices of the beamforming transformation vectors given by $\overline{\bfA} = [\overline{\bfa}_1, \cdots, \overline{\bfa}_K]$, where $\overline{\bfa}_{\kc}$ is the optimal beamforming solution for the $\kc$-th choice of the user index and it contains both the common and the private precoders stacked together as $\overline{\bfa}_{\kc} = [\bfa_{\ct}(\kc)^{\Tm}, \bfa_{1}(\kc)^{\Tm}, \cdots, \bfa_{K}(\kc)^{\Tm}]^{\Tm}$. After this, the user index $k_{\text{opt}}$ which maximises the sum-rate is selected, i.e., 
\begin{align}
k_{\text{opt}} = \max\limits_{\kc} \Bigg[\sum\limits_{i=1}^{K}& \log_2\left(1 + \gamma_{\text{p},i}^{\mathrm{lb}}(\overline{\bfa}_{\kc})  \right) \: +  \nonumber \\
 &\min\limits_{j}\:\log_2\left(1 + \gamma_{\ct,j}^{\mathrm{lb}}(\overline{\bfa}_{\kc})\right) \Bigg]. \label{eqn11}
\end{align}
The objective function in \eqref{eqn10} is still very involved, which can be further simplified by the fractional programming~(FP) approach of \cite{fractional1, fractional2} to a more tractable form as follows
\begin{align}
    &\log\left(1 + \lambda_{\text{c},\kc} \right) - \lambda_{\text{c},\kc}  + 2\left(\sqrt{1 + \lambda_{\text{c},\kc}}\right)\text{Re}\left\{{{\beta}}_{\text{c},\kc}^{*}\bfz_{\kc}^{\Hm}\bfa_{\text{c}}\right\} \nonumber \\
    &- |{{\beta}}_{\text{c},\kc}|^2 \Bigg(\left|{\bfz_{\kc}^{\Hm}\bfa_{\ct}}\right|^2 + \bfa_{\ct}^{\Hm}\bfZ_{\kc}\bfa_{\ct} 
 + \left|{\bfq_{\kc}^{\Hm}\bfa_{\kc}}\right|^2  \nonumber \\
 & \quad \quad \quad  +\sum\limits_{j=1}^{K}{\bfa_j^{\Hm}\bfQ_{j,\kc} \bfa_j} + 1 \Bigg) \nonumber \\
    &+ \sum\limits_{i=1}^{K}\left(\log\left(1 + \lambda_i \right) - \lambda_i + 2\left(\sqrt{1 + \lambda_i}\right)\text{Re}\left\{{{\beta}}_i^{*}\bfq_i^{\Hm}\bfa_i\right\} \right) \nonumber \\& - \sum\limits_{i=1}^{K}|{{\beta}}_i|^{2} \left( \left|{\bfq_i^{\Hm}\bfa_i}\right|^2 + \sum\limits_{j=1}^{K}\bfa_j^{\Hm}\bfQ_{j,i} \bfa_j + 1 \right)  \label{fp1}
\end{align}
where $\lambda_i$, $\lambda_{\text{c},i}$ , ${{\beta}}_i $ and ${\beta}_{\ct,i} $ are the auxiliary variables of the FP approach. Problem \eqref{P1} is heuristically optimised by solving problem \eqref{eqn10} for each $k_{\ct} \in \{1, \cdots, K \}$ choice of user index, which is done by maximising the reformulated function in \eqref{fp1} using the iterative non-convex block-coordinate descent~(BCD) method \cite{bcd, bcdmain}, where the auxiliary variables and the transformation vectors are alternatingly updated in each iteration as discussed in the next section.
\section{Solution Approach using BCD Method}
Keeping $\bfa_{\ct}$ and $\bfa_k$ fixed, the maximisation of \eqref{fp1} with respect to each of the auxiliary variables (keeping the other auxiliary variables fixed) is a convex optimisation problem. The closed-form solution of the auxiliary variables alternatingly maximising the objective function in \eqref{fp1} reads as \cite[Sec. IV]{syed2024design}
\begin{align}
    \lambda_i = \gamma_{\text{p},i}^{\mathrm{lb}},   \quad \lambda_{\text{c},k} = \gamma_{\ct,k}^{\mathrm{lb}} \label{lamk}
\end{align}
%{ \small
\begin{align}
   {{\beta}}_i =  \dfrac{\left(\sqrt{1 + \lambda_i}\right)\bfq_i^{\Hm}\bfa_i}{\left|{\bfq_i^{\Hm}\bfa_i}\right|^2 + \sum\nolimits_{j=1}^{K}{\bfa_j^{\Hm}\bfQ_{j,i} \bfa_j} + 1}   \label{betak1}
   \end{align}
   \begin{align}
   {{\beta}}_{\text{c},k} =  \dfrac{\left(\sqrt{1 + \lambda_{\ct,k}}\right)\bfz_k^{\Hm}\bfa_{\ct}}{\left|{\bfz_k^{\Hm}\bfa_{\ct}}\right|^2 + \bfa_{\ct}^{\Hm}\bfZ_k\bfa_{\ct} 
 + \left|{\bfq_k^{\Hm}\bfa_k}\right|^2 + \sum\limits_{j=1}^{K}{\bfa_j^{\Hm}\bfQ_{j,k} \bfa_j} + 1}.  \label{betak}
\end{align}
For updating the beamforming transformation matrices, we firstly denote $\bfv = [\bfa_{\ct}^{\Tm}, \bfa_1^{\Tm}, \cdots, \bfa_K^{\Tm}]^{\Tm} \in \mathbb{C}^{2 M K T_{\text{dl}}}$, and introduce selection matrices $\bfD_\text{c} \in \mathbb{C}^{K M T_{\text{dl}}  \times 2 K M T_{\text{dl}} }, \: \text{and} \: \bfD_1,\cdots, \bfD_K \in \mathbb{C}^{ M T_{\text{dl}}  \times 2 K M T_{\text{dl}} }$ such that  $\bfa_\text{c} = \bfD_\text{c} \bfv$ and $\bfa_i = \bfD_i \bfv$. The selection matrices are given by
\begin{align*}
    &\bfD_\text{c} = \left[{\mathbf{I}}_{K M T_{\text{dl}}},  {\mathbf{0}}_{ K M T_{\text{dl}} \times  K M T_{\text{dl}}}\right] \\
    &\bfD_i = [\underbrace{{\mathbf{0}}_{M T_{\text{dl}} \times M T_{\text{dl}}},\cdots}_{(K + i -1).M T_{\text{dl}}}, {\mathbf{I}}_{M T_{\text{dl}}},  \cdots, {\mathbf{0}}_{M T_{\text{dl}} \times M T_{\text{dl}}}].  
\end{align*} 
The DL power constraint can now be written as $\bfv^{\Hm} \bfS \bfv \leq P_{\text{t}}$, where $\bfS = \bfD_{\ct}^{\Hm} \bfF \bfD_{\ct} + \sum\nolimits_{i=1}^{K} \bfD_i^{\Hm} \bfF_i \bfD_i$. Optimisation of $\bfv$ for each common rate user $\kc$ is also a convex optimisation problem, however, the DL power constraint requires a one-dimensional search of the optimal dual variable corresponding to the inequality constraint \eqref{eqn9}. This increases the complexity of the update step of the transformation matrices. This problem can be easily circumvented by applying the rescaling trick mentioned in \cite{joham2002transmit, christensen2008weighted, zhao2023rethinking}, where the inequality constraint is incorporated in the objective function by exploiting the fact that the inequality constraint in \eqref{eqn9} will be satisfied with equality for the optimal $\bfv$ (see \cite[Sec. IV]{syed2024design}). With $\bfv^{\Hm}\bfS \bfv = P_{\text{t}}$, the objective function in \eqref{fp1} can be reformulated for the update of $\bfv$ as
\begin{align}
    & 2\sqrt{1 + \lambda_{\text{c},\kc}}\text{Re}\left\{{{\beta}}_{\text{c},\kc}^{*}\bfz_{\kc}^{\Hm}\bfD_{\ct} \bfv\right\}  - |{{\beta}}_{\text{c},\kc}|^2 \left|{\bfz_{\kc}^{\Hm}\bfD_{\ct} \bfv}\right|^2 -  |{{\beta}}_{\text{c},\kc}|^2 \nonumber \\
    & \times \Big(\bfv ^{\Hm}\bfD_{\ct}^{\Hm}\bfZ_{\kc}\bfD_{\ct} \bfv   + \left|{\bfq_{\kc}^{\Hm}\bfD_{\kc} \bfv}\right|^2 + \sum\limits_{j=1}^{K}{\bfv^{\Hm}\bfD_j^{\Hm}\bfQ_{j,\kc} \bfD_j \bfv} \Big) \nonumber \\
    &- |{{\beta}}_{\text{c},\kc}|^2 \dfrac{\bfv^{\Hm}\bfS \bfv}{P_{\text{t}}} + \sum\limits_{i=1}^{K} 2\left(\sqrt{1 + \lambda_i}\right)\text{Re}\left\{{{\beta}}_i^{*}\bfq_i^{\Hm}\bfD_i \bfv\right\}  \nonumber \\& - \sum\limits_{i=1}^{K}|{{\beta}}_i|^{2} \left( \left|{\bfq_i^{\Hm}\bfD_i \bfv}\right|^2 + \sum\limits_{j=1}^{K}\bfv^{\Hm}\bfD_j^{\Hm}\bfQ_{j,i} \bfD_j \bfv + \dfrac{\bfv^{\Hm}\bfS \bfv}{P_{\text{t}}} \right)  \label{fp2}
\end{align}
where we have ignored the terms independent of $\bfv$. The optimal solution of $\bfv$ depending on the index $\kc$ can be obtained as
\begin{align}
    \bfv'({\kc}) = \bfY_{\kc}^{-1} \bfx_{\kc}  \label{v}
\end{align}
\begin{align}
   &\quad \bfY_{\kc} =  \sum\limits_{i=1}^{K}|{{\beta}}_i|^{2}  \left( {\bfD_i^{\Hm}\bfq_i\bfq_i^{\Hm}\bfD_i } + \sum\limits_{j=1}^{K}\bfD_j^{\Hm}\bfQ_{j,i} \bfD_j  + \dfrac{\bfS}{P_{\text{t}}} \right) \nonumber \\ & \quad+ |{{\beta}}_{\text{c},\kc}|^{2} \Bigg(\bfD_{\ct}^{\Hm}\bfZ_{\kc}\bfD_{\ct}  + {\bfD_{\kc}^{\Hm}\bfq_{\kc}\bfq_{\kc}^{\Hm}\bfD_{\kc} }  \Bigg)  \\
   & \quad + |{{\beta}}_{\text{c},\kc}|^{2} \left( \sum\limits_{j=1}^{K}{\bfD_j^{\Hm}\bfQ_{j,\kc} \bfD_j } + \bfD_{\ct}^{\Hm}\bfz_{\kc} \bfz_{\kc}^{\Hm}\bfD_{\ct} + \dfrac{\bfS}{P_{\text{t}}}\right) \nonumber
   \end{align}
\begin{align}
   & \quad \bfx_{\kc}^{\Hm} = \sum\limits_{i=1}^{K} \left(\sqrt{1 + \lambda_i}\right){{\beta}}_i^{*}\bfq_i^{\Hm}\bfD_i  +  (\sqrt{1 + \lambda_{\text{c},\kc}}){{\beta}}_{\ct,\kc}^{*}\bfz_{\kc}^{\Hm}\bfD_{\ct}.   \nonumber
\end{align}   
\par Even though the DL power constraint has been incorporated into the objective function of \eqref{fp2}, the solution \eqref{v} must be rescaled to ensure that the constraint is satisfied with equality (cf.~\cite{joham2002transmit, christensen2008weighted, zhao2023rethinking}). Therefore, the rescaled solution is given by 
\begin{align}
    \bfv(\kc) = \dfrac{\sqrt{P_{\text{t}}}}{\sqrt{\bfv'^{\Hm}({\kc}) \bfS \bfv'({\kc})}}\bfv'({\kc}).  \label{v2}
\end{align}
Finally, after obtaining the optimal $\bfv(\kc)$ for each choice of $\kc$ by \eqref{v2}, we can obtain the optimal user index $k_{\text{opt}}$ by solving \eqref{eqn11}, thereby yielding 
\begin{align}
    \bfv_{\text{opt}} =  \bfv(k_{\text{opt}}).  \label{eqn20}
\end{align}
\eqref{P1} is thus solved suboptimally in closed-form by replacing its objective function by \eqref{fp1} and alternatingly applying \eqref{lamk}, \eqref{betak1}, \eqref{betak}, \eqref{v} and \eqref{eqn20} till the convergence is reached. Note that the precoders obtained by \eqref{eqn20} only depend on the second-order channel statistics and the pilot matrix, hence requiring optimisation only in the covariance matrix coherence interval. The overall RS BP algorithm is summarised in Algorithm 1.
\begin{algorithm}
\caption{Rate Splitting with Bilinear Precoders (RS BP)}
\begin{algorithmic}[1]
    \State  Initialise $\bfv$ by setting $\bfa_{\ct}$ as $\sum\nolimits_k\bfz_k$, $\bfa_k$ as $\bfq_k$ and stacking them together in a vector, and then rescale the vector to satisfy $\bfv^{\Hm}\bfS\bfv = P_{\text{t}}$;    
    \State {\textbf{Repeat}}
        \State Update $\lambda_i$, $\lambda_{\ct,k}$, $\beta_i$ and $\beta_{\ct,k}$ according to \eqref{lamk}, \eqref{betak1}, and \eqref{betak} respectively;
        \State Find $\bfv (\kc)$ $\forall$ $\kc \in \{1, \cdots, K \}$ and compute $\overline{\bfA}$;
        \State Determine $k_{\text{opt}}$ by \eqref{eqn11} and then compute $\bfv_{\text{opt}}$ by \eqref{eqn20};  
        \If{the objective value in \eqref{P1} decreases} 
            \State break;
        \Else
           \State $\bfa_{\ct} = \bfD_{\ct} \bfv_{\text{opt}}$, $\bfa_k = \bfD_k \bfv_{\text{opt}}$ ;
        \EndIf
    \State \textbf{Until} The value of the objective function in \eqref{P1} converges.      
\end{algorithmic}
\end{algorithm}
\section{Complexity of the Algorithm}
The complexity of the proposed RS BP algorithm is dominated by computing $K$ times the inverse of the matrix $\bfY_k$ in every iteration. If $I_{\text{iter}}$ denotes the total number of iterations required for the convergence of RS BP algorithm, its complexity is given by $\Ocal \left(I_{\text{iter}}K \left(2 M K T_{\text{dl}} \right)^3 \right)$. However, note that the entire process needs to be performed just once in every coherence interval of the covariance matrices, i.e., we do not need to update $\bfa_{\ct}$ and $\bfa_k$ whenever the channels change. On the other hand, the complexity of the SINR increasing algorithm of \cite{amor2023rate} is given by $\Ocal \left(2\:I_0 I_1 K \left(M K T_{\text{dl}} \right)^3 + 2 I_0 I_2 K \left( M T_{\text{dl}} \right)^3 \right)$. Here, $I_1$ denotes the number of iterations required for the common precoder optimisation and $I_2$ is the number of iterations required to compute the optimal dual variable for the DL power constraint. $I_0$ is the number of iterations required for obtaining the optimal power allocation and the factor 2 is due to the one-dimensional golden section method employed to optimise the power split. In order to have a better intuition of the complexity of the two algorithms, we compare the average run-times in Table~\ref{tab:table1} for $T_{\text{dl}} = 4$ and $P = 30$ dB. It is evident from Table~\ref{tab:table1} that the average run-time of the proposed RS BP algorithm is much less (more than a factor of 10) than that of the algorithm in \cite{amor2023rate}. We also compare the run-time complexity with that of the stochastic iterative weighted minimum mean squared error~(SIWMMSE) algorithm of \cite{bruno}, where the common and private precoders are optimised in every channel coherence interval using the imperfect CSI available at the BS. The computational complexity of this algorithm is very high as it requires the entire optimisation process to be performed in every channel coherence interval. Additionally, unlike the RS BP and the SINR increasing \cite{amor2023rate} algorithms, the SIWMMSE approach does not have closed-form update steps and it uses the CVX interior-point solver \cite{grant2014cvx}, making its complexity even higher. 
\begin{table}[h!]
\renewcommand{\arraystretch}{1.5} % Adjust this value to control row spacing
  \begin{center}
   \scriptsize %\tiny %
    \caption{Complexity Comparison}
    \label{tab:table1}
    \begin{tabular}{|p{1.3cm}|p{1.4cm}|p{2.58cm}|p{0.8cm}|} % <-- Alignments: 1st column left, 2nd middle and 3rd right, with vertical lines in between
    \hline
      \textbf{Algorithm} & \textbf{Frequency of Optimisation } & \textbf{Complexity} & \textbf{Run-Time [sec]}\\
      \hline
      \textbf{RS BP} & \textbf{Covariance matrix} coherence int.  & $\Ocal \Big(I_{\text{iter}}K (2 M K T_{\text{dl}})^3 \Big)$ & \textbf{0.67}\\
      \hline
      SINR Increasing Algorithm \cite{amor2023rate} & \textbf{Covariance matrix } coherence int. & $\Ocal \Big(2 I_0 I_1 K(M K T_{\text{dl}})^3  +2 I_0 I_2 K( M T_{\text{dl}})^3 \Big)$ & 11.56\\
      \hline
      SIWMMSE \cite{bruno} & \textbf{CSI} coherence~int. & Uses CVX & 24.09 \\ 
     \hline
    \end{tabular}
  \end{center}
\end{table}
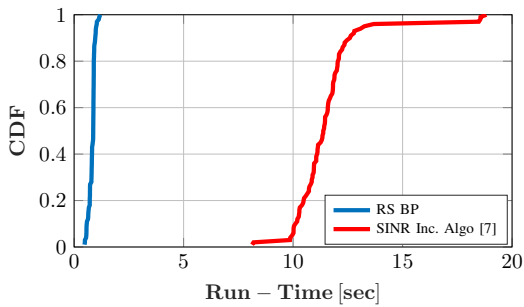
\begin{figure}		
		\scalebox{0.80}{\input{run_time}} %{\input{Convergence}} %{\input{conv_cdf}} %
		\caption{CDF plot of the run-time needed until convergence for $M= 16, K = 4, T_{\text{dl}} = 4$ at $P_{\text{t}} = 30$ dB}
		\label{convergence}
\end{figure}
\section{Results}
In this section, numerical results are provided to validate the effectiveness of the proposed algorithm. The system consists of one BS equipped with $M = 16$ antennas serving $K = 4$ single-antenna users with varying number of training pilots. We consider the cases where the number of pilots used for channel estimation is less than that required
for a contamination-free estimation, i.e., $T_{\text{dl}} < M$. The covariance matrices $\bfC_k$ are generated according to the urban micro channel model described in the 3GPP technical report~\cite{etsi5138}. The details of the covariance matrix generation are provided in \cite[Sec. V]{syed2024design}. We firstly compare the computational complexity of the proposed RS BP algorithm and that of the SINR Increasing algorithm \cite{amor2023rate} in Fig.~~\ref{convergence}, where the cumulative distribution function (CDF) of the run-time needed until convergence is plotted for $T_{\text{dl}} = 4$ at $P_{\text{t}} = 30$ dB. It is evident from the figure that the run-time of the proposed RS BP algorithm is negligible compared to that of \cite{amor2023rate}. 
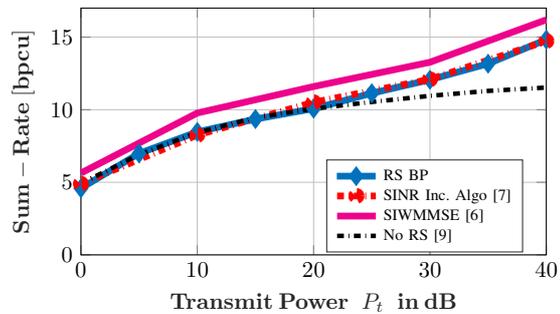
\begin{figure}		
		\scalebox{0.85}{\input{Half_pilot_hs}} 
		\caption{Achievable Sum-Rate vs $P_{\text{t}}$ for $M= 16, K~=~4, T_{\text{dl}} = 8$}
		\label{half}
\end{figure}
\begin{figure}		
		\scalebox{0.85}{\input{Quart_pilot_hs}} 
		\caption{Achievable Sum-Rate vs $P_{\text{t}}$ for $M= 16, K~=~4, T_{\text{dl}} = 4$}
		\label{quarter}
\end{figure}
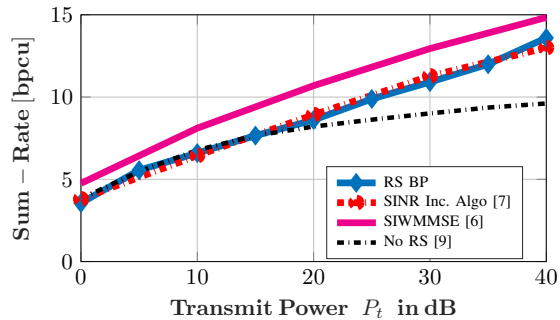

\begin{figure}		
		\scalebox{0.85}{\input{Less_pilot_hs}}
		\caption{Achievable Sum-Rate vs $P_{\text{t}}$ for $M= 16, K~=~4, T_{\text{dl}} = 2$}
		\label{less}
\end{figure}
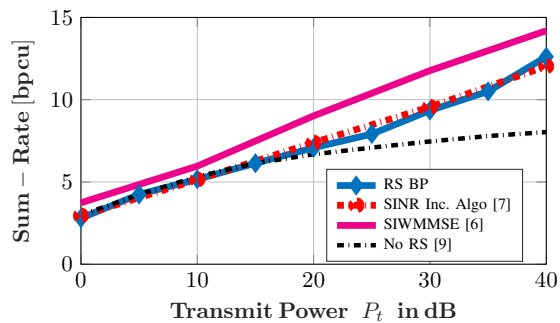
\par The average achievable instantaneous sum-rate of the users is taken as the performance metric in our analysis, which is computed with the different algorithms and compared over the DL transmit power levels $P_{\text{t}}$. The estimation error variance $\sigma_k^2$ is set to 1 for all the users. The rate is averaged over 100 covariance matrices, which are generated by varying the distance between the BS and the users and the path loss factors of the scatterers (cf.~\cite[Sec. V]{syed2024design}). For each of the generated covariance matrices, the sum-rate is averaged over 500 channel realisations. The performance of the proposed RS BP algorithm is compared with the following algorithms: (i) SINR Increasing algorithm of \cite{amor2023rate}, (ii) SIWMMSE algorithm of \cite{bruno}, and (iii) BP based non-RS approach of \cite{amor2020bilinear}.
 Figs.~\ref{half}, ~\ref{quarter} and ~\ref{less} compare the average achievable sum-rate of the users for the different schemes with respect to the transmit power $P_{\text{t}}$ in dB for $T_{\text{dl}} = 8, 4$ and 2 respectively. It has been shown in \cite{amor2020bilinear} and \cite{syed2024design} that the BP based approach leads to a very cheap online data transmission phase, where the pre-computed transformation matrices are multiplied with the channel observations $\bfy_k$ available at the BS in every CSI coherence interval. But it comes at a cost of sum-rate saturation in the high power regime. It can be observed from Figs.~\ref{half} and~\ref{quarter}, where we have the case of $T_{\text{dl}} \geq K$, the sum-rate starts saturating at high $P_{\text{t}}$ for the non-RS setup. This is attributed to the fact that the BP cannot mitigate the inter-user interference in the presence of imperfect CSI and $T_{\text{dl}} < M$ training pilots. It can also be observed that the sum-rate starts growing asymptotically by employing the RS algorithms. Moreover, the propsed RS BP algorithm performs almost similar to the SINR increasing algorithm, despite being computationally much cheaper. As expected, the SIWMMSE based RS algorithm of \cite{bruno} performs the best among other methods, but it is to be noted that this algorithm requires the optimisation process (which involves CVX) to be performed in every channel coherence interval, which is practically infeasible due to huge computational complexity. In Fig.~\ref{less}, where we have the case of $T_{\text{dl}} < K$ pilots, it can still be observed that employing RS leads to an asymptotic increase of the achievable sum-rate at high transmit power.
\section{Conclusion}
In this work, we have presented a low-complexity RS strategy for the multi-user MISO systems operating in FDD mode based on bilinear precoders. The simulation results illustrate that a performance gain can be achieved by employing the proposed algorithm even in the presence of imperfect CSI and even when the number of training pilots is less than the number of users. The optimisation process only depends on the second-order channel statistics which needs to be performed only once in every $\textbf{covariance matrix coherence interval}$, which significantly reduces the computational complexity. 
\section{Appendix}
\subsection{Closed-Form SINR Expressions}
Denoting $\bfh = \left[\bfh_1^{\Tm}, \cdots, \bfh_K^{\Tm}   \right]^{\Tm}$ and $\bfn = \left[\bfn_1^{\Tm}, \cdots, \bfn_K^{\Tm}   \right]^{\Tm}$, we can write
\begin{align*}
    \bfh_k = {\mathbf{E}}_k\bfh, \quad \bfn_k = {\mathbf{E}}_k\bfn, \quad \bfy = \bfD \bfh + \bfn, \quad \bfC_{\bfh} = \Ex [ \bfh \bfh^{\Hm}] 
\end{align*}
where ${\mathbf{E}}_k = \bfe_k^{\Tm} \otimes {\mathbf{I}}_M$ and $\bfD = {\mathbf{I}}_K \otimes \Phimat^{\Hm} $.
We can compute the closed-form expressions of the terms in \eqref{lb_c} and \eqref{lb_p} as done in \cite[Appendix]{amor2023rate} 
\begin{align}
&\mathop{{}\mathbb{E}}[\bfh_k^{\Hm}\bfp_{\ct}] = \bfz_k^{\Hm}\bfa_{\ct} \\
&\mathrm{var}({\bfh_k^{\Hm}\bfp_{\ct}})  = \mathop{{}\mathbb{E}}[|{{\bfh_k^{\Hm}}{\bfp}_{\ct}} -\mathop{{}\mathbb{E}}[\bfh_k^{\Hm}\bfp_{\ct}]|^2 ] = \bfa_{\ct}^{\Hm}\bfZ_k\bfa_{\ct} \\
&\mathop{{}\mathbb{E}}[\bfh_k^{\Hm}\bfp_k] = \bfq_k^{\Hm}\bfa_k, \quad  \Ex[|{{\bfh_k^{\Hm}}{\bfp_j}}|^2] = \bfa_j^{\Hm}\bfQ_{j,k} \bfa_j \\
&\mathrm{var}({\bfh_k^{\Hm}\bfp_{k}})  = \mathop{{}\mathbb{E}}[|{{\bfh_k^{\Hm}}{\bfp}_{k}} -\mathop{{}\mathbb{E}}[\bfh_k^{\Hm}\bfp_{k}] |^2] = \bfa_k^{\Hm}\bfQ_{k,k} \bfa_k 
\end{align}
where we have
\begin{align*}
  &\bfz_k = \left(\bfD^{*} \otimes \bfE_k \right)\bfc_{\bfh} , \quad \bfc_{\bfh} = \text{vec}(\bfC_{\bfh}), \quad \bfC = \Ex \left[ \bfh \bfh^{\Hm}\right], \\
  &\bfZ_k = \bfC_{\bfy}^{\Tm} \otimes \bfC_k, \quad \bfc_k = \text{vec}(\bfC_k), \quad \bfq_k = \left(\Phimat^{\Tm} \otimes {\mathbf{I}}_M \right) \bfc_k, \\
  &\bfQ_{i,k} = \bfC_{\bfy_i}^{\Tm} \otimes \bfC_k, \quad \bfC_{\bfy} = \text{blkdiag}(\bfC_{\bfy_1}, \cdots, \bfC_{\bfy_K}).
\end{align*}

\bibliographystyle{IEEEtran}
\bibliography{bibliography}
\end{document}

%% file: macros.tex
\newcommand{\bfa}{{\boldsymbol a}}

\newcommand{\bfc}{{\boldsymbol c}}

\newcommand{\bfe}{{\boldsymbol e}}

\newcommand{\bfh}{{\boldsymbol h}}

\newcommand{\bfn}{{\boldsymbol n}}

\newcommand{\bfp}{{\boldsymbol p}}
\newcommand{\bfq}{{\boldsymbol q}}

\newcommand{\bfs}{{\boldsymbol s}}

\newcommand{\bfv}{{\boldsymbol v}}

\newcommand{\bfx}{{\boldsymbol x}}
\newcommand{\bfy}{{\boldsymbol y}}
\newcommand{\bfz}{{\boldsymbol z}}
\newcommand{\bfA}{{\boldsymbol A}} %\newcommand{\bfA}{{\mathbf A}}

\newcommand{\bfC}{{\boldsymbol C}}
\newcommand{\bfD}{{\boldsymbol D}}
\newcommand{\bfE}{{\boldsymbol E}}
\newcommand{\bfF}{{\boldsymbol F}}

\newcommand{\bfQ}{{\boldsymbol Q}}

\newcommand{\bfS}{{\boldsymbol S}}

\newcommand{\bfY}{{\boldsymbol Y}}
\newcommand{\bfZ}{{\boldsymbol Z}}

\newcommand{\Phimat}{{\boldsymbol{\it{\Phi}}}} %\newcommand{\Phimat}{\mathbf{\Phi}}

\newcommand{\Tm}{\mathrm{T}}

\newcommand{\nc}{\mathcal{N_\mathbb{C}}}

\newcommand{\Hm}{{\mathrm{H}}}

\newcommand{\Ex}{\mathop{{}\mathbb{E}}}

\newcommand{\Ocal}{\mathcal{O}}
\newcommand{\ct}{\text{c}}
\newcommand{\kc}{k_{\text{c}}}

%% file: run_time.tex
\definecolor{mycolor1}{rgb}{0.00000,0.44700,0.74100}%
\definecolor{mycolor2}{rgb}{0.85000,0.32500,0.09800}%
\begin{tikzpicture}

\begin{axis}[%
width=0.9\figW,
height=\figH,
at={(0\figW,0\figH)},
scale only axis,
xmin=0,
xmax=20,
xlabel style={font=\color{white!15!black}},
xlabel={$\bf{Run-Time \:[sec]}$},
ymin=0,
ymax=1,
ylabel style={font=\color{white!15!black}},
ylabel={$\bf{CDF}$},
axis background/.style={fill=white},
xmajorgrids,
ymajorgrids,
legend style={at={(0.99,0.01)}, anchor=south east, legend cell align=left, align=left, draw=white!15!black, row sep=-0.05cm, font = \scriptsize}
]
\addplot [color=mycolor1, solid, line width=2.0pt]
  table[row sep=crcr]{%
0.476104	0.01\\
0.4917482	0.02\\
0.4935561	0.03\\
0.5468334	0.04\\
0.5573353	0.05\\
0.5620364	0.06\\
0.5643025	0.07\\
0.5659517	0.08\\
0.5685004	0.09\\
0.577956	0.1\\
0.5874526	0.11\\
0.633785	0.12\\
0.6445545	0.13\\
0.6457453	0.14\\
0.6531969	0.15\\
0.6633618	0.16\\
0.6742564	0.17\\
0.7136826	0.18\\
0.7138393	0.19\\
0.7151597	0.2\\
0.715451	0.21\\
0.7227947	0.22\\
0.7229331	0.23\\
0.7240443	0.24\\
0.7254924	0.25\\
0.7255029	0.26\\
0.7316281	0.27\\
0.7894403	0.28\\
0.7897282	0.29\\
0.7941299	0.3\\
0.7961381	0.31\\
0.7997695	0.32\\
0.800577	0.33\\
0.802594	0.34\\
0.8063817	0.35\\
0.8084082	0.36\\
0.8094044	0.37\\
0.8106068	0.38\\
0.8146659	0.39\\
0.8162895	0.4\\
0.8182746	0.41\\
0.8209842	0.42\\
0.821081	0.43\\
0.8397068	0.44\\
0.8522616	0.45\\
0.8661423	0.46\\
0.8697921	0.47\\
0.869869	0.48\\
0.8716063	0.49\\
0.8719954	0.5\\
0.8721414	0.51\\
0.8738746	0.52\\
0.8754302	0.53\\
0.8756477	0.54\\
0.8769593	0.55\\
0.8781196	0.56\\
0.880103	0.57\\
0.8802888	0.58\\
0.8819504	0.59\\
0.8823432	0.6\\
0.8825338	0.61\\
0.8831471	0.62\\
0.8832073	0.63\\
0.8861202	0.64\\
0.8861611	0.65\\
0.8877692	0.66\\
0.8879718	0.67\\
0.8896509	0.68\\
0.8923112	0.69\\
0.8933261	0.7\\
0.8935773	0.71\\
0.8940834	0.72\\
0.8961288	0.73\\
0.8976322	0.74\\
0.8977881	0.75\\
0.8980555	0.76\\
0.8985919	0.77\\
0.8986006	0.78\\
0.8988677	0.79\\
0.8989058	0.8\\
0.9032763	0.81\\
0.9089563	0.82\\
0.9093342	0.83\\
0.9156601	0.84\\
0.9175402	0.85\\
0.9178948	0.86\\
0.9261296	0.87\\
0.9456928	0.88\\
0.9566741	0.89\\
0.9643418	0.9\\
0.9734666	0.91\\
0.9855325	0.92\\
0.9877711	0.93\\
0.9932003	0.94\\
1.0114584	0.95\\
1.0350324	0.96\\
1.0402866	0.97\\
1.1267586	0.98\\
1.1497868	0.99\\
1.1950082	1\\
};
\addlegendentry{RS BP}

\addplot [color=red, solid, line width=2.0pt]
  table[row sep=crcr]{%
8.1245865	0.01\\
8.1951052	0.02\\
9.856497	0.03\\
9.8700284	0.04\\
9.9569439	0.05\\
10.0168188	0.06\\
10.0217348	0.07\\
10.0228394	0.08\\
10.0389849	0.09\\
10.1036344	0.1\\
10.187596	0.11\\
10.1952504	0.12\\
10.2579984	0.13\\
10.2705088	0.14\\
10.2855537	0.15\\
10.287749	0.16\\
10.3069779	0.17\\
10.4051856	0.18\\
10.4613175	0.19\\
10.4793726	0.2\\
10.4992033	0.21\\
10.583285	0.22\\
10.6523165	0.23\\
10.7224729	0.24\\
10.7227616	0.25\\
10.7459781	0.26\\
10.8067058	0.27\\
10.850823	0.28\\
10.8524952	0.29\\
10.8883922	0.3\\
10.8887259	0.31\\
10.938267	0.32\\
10.9465949	0.33\\
10.9498455	0.34\\
10.9542174	0.35\\
10.9614643	0.36\\
11.0555973	0.37\\
11.0566522	0.38\\
11.0576659	0.39\\
11.0979572	0.4\\
11.1153226	0.41\\
11.1161537	0.42\\
11.1331128	0.43\\
11.1411245	0.44\\
11.2558619	0.45\\
11.3085018	0.46\\
11.3363048	0.47\\
11.3434278	0.48\\
11.3762089	0.49\\
11.3767452	0.5\\
11.4134781	0.51\\
11.4204223	0.52\\
11.4466835	0.53\\
11.4573067	0.54\\
11.4589824	0.55\\
11.4628428	0.56\\
11.5306818	0.57\\
11.5655122	0.58\\
11.5835069	0.59\\
11.583842	0.6\\
11.5942452	0.61\\
11.6064946	0.62\\
11.6303498	0.63\\
11.6824617	0.64\\
11.7220282	0.65\\
11.7882107	0.66\\
11.8086148	0.67\\
11.8177556	0.68\\
11.822331	0.69\\
11.8225609	0.7\\
11.8343315	0.71\\
11.8773899	0.72\\
11.8887344	0.73\\
11.8919301	0.74\\
11.9275899	0.75\\
11.9972458	0.76\\
12.0095408	0.77\\
12.0291253	0.78\\
12.0306552	0.79\\
12.0450889	0.8\\
12.0883477	0.81\\
12.0926172	0.82\\
12.0975358	0.83\\
12.1517285	0.84\\
12.2386778	0.85\\
12.2633809	0.86\\
12.3428541	0.87\\
12.3565665	0.88\\
12.4557393	0.89\\
12.5964573	0.9\\
12.6355349	0.91\\
12.7720569	0.92\\
12.7982149	0.93\\
13.0807929	0.94\\
13.2493729	0.95\\
13.6778182	0.96\\
18.4812648	0.97\\
18.5509123	0.98\\
18.5600176	0.99\\
18.8274703	1\\
};
\addlegendentry{SINR Inc. Algo \cite{amor2023rate}}
\end{axis}
\end{tikzpicture}%

%% file: Half_pilot_hs.tex
\definecolor{mycolor1}{rgb}{0.00000,0.44700,0.74100}%
\definecolor{mycolor2}{rgb}{0.85000,0.32500,0.09800}%
\begin{tikzpicture}

\begin{axis}[%
width=0.9\figW,
height=\figH,
at={(0\figW,0\figH)},
scale only axis,
xmin=0,
xmax=40,
xlabel style={font=\color{white!15!black}},
xlabel={$\bf{Transmit\:Power\:}$ $P_{t}$ $\bf{\:in\:dB}$},
ymin=0,
ymax=17,
ylabel style={font=\color{white!15!black}},
ylabel={$\bf{Sum-Rate\:[bpcu]}$},
axis background/.style={fill=white},
xmajorgrids,
ymajorgrids,
legend style={at={(0.95,0.01)}, anchor=south east, legend cell align=left, align=left, draw=white!15!black, row sep=-0.05cm, font=\scriptsize}
]
\addplot [color=mycolor1, solid, line width=3.0pt, mark=diamond, mark options={solid, mycolor1}]
  table[row sep=crcr]{%
0	4.58815298253579\\
5	6.9878030812724\\
10	8.47040864567478\\
15	9.35455977903047\\
20	10.0714139522121\\
25	11.1347710985671\\
30	12.063236467544\\
35	13.1692738131282\\
40	14.8279021206393\\
};
\addlegendentry{RS BP}

\addplot [color=red, dashdotted, line width=3.0pt, mark=o, mark options={dashdotted, red}]
  table[row sep=crcr]{%
0	4.89497224250204\\
10	8.25660602792495\\
20	10.5118484818251\\
30	12.093236467544\\ 
40	14.7279873500166\\
};
\addlegendentry{SINR Inc. Algo \cite{amor2023rate}}

\addplot [color=magenta, solid, line width=3.0pt, mark=None, mark options={solid, magenta}]
  table[row sep=crcr]{%
0	5.63490719376304\\
10	9.771010137377560\\ 
20	11.596292349842562\\
30	13.269495752867819\\ 
40	16.19436912850571\\
};
\addlegendentry{SIWMMSE \cite{bruno}}

\addplot [color=black, dashdotted, line width=2.0pt, mark=none, mark options={dashdotted, black}]
  table[row sep=crcr]{%
0	4.91015546083617\\
5	6.90693915706021\\
10	8.42418450415506\\
15	9.44666635298829\\
20	10.0714139522121\\
25	10.5347710985671\\
30	10.9553237221165\\
35	11.2946146179173\\
40	11.5266759274894\\
};
\addlegendentry{No RS \cite{amor2020bilinear}}

\end{axis}

\end{tikzpicture}%

%% file: Quart_pilot_hs.tex
\definecolor{mycolor1}{rgb}{0.00000,0.44700,0.74100}%
\definecolor{mycolor2}{rgb}{0.85000,0.32500,0.09800}%
\begin{tikzpicture}

\begin{axis}[%
width=0.9\figW,
height=\figH,
at={(0\figW,0\figH)},
scale only axis,
xmin=0,
xmax=40,
xlabel style={font=\color{white!15!black}},
xlabel={$\bf{Transmit\:Power\:}$ $P_{t}$ $\bf{\:in\:dB}$},
ymin=0,
ymax=15,
ylabel style={font=\color{white!15!black}},
ylabel={$\bf{Sum-Rate\:[bpcu]}$},
axis background/.style={fill=white},
xmajorgrids,
ymajorgrids,
legend style={at={(0.95,0.01)}, anchor=south east, legend cell align=left, align=left, draw=white!15!black, row sep=-0.05cm, font=\scriptsize}
]
\addplot [color=mycolor1, solid, line width=3.0pt, mark=diamond, mark options={solid, mycolor1}] table[row sep=crcr]{%
0	3.56023974939997\\
5	5.5536889104674\\
10	6.59361916084178\\
15	7.65574653148277\\
20	8.60674833973187\\
25	9.88356406714752\\
30	10.9103525205014\\
35	11.9927467289041\\
40	13.5925254364015\\
};
\addlegendentry{RS BP}

\addplot [color=red, dashdotted, line width=3.0pt, mark=o, mark options={dashdotted, red}]
  table[row sep=crcr]{%
0	3.80663438907286\\
10	6.43895262373829\\
20	8.92847020281144\\
30	11.2821146323846\\
40	12.9950316033764\\
};
\addlegendentry{SINR Inc. Algo \cite{amor2023rate}}

\addplot [color=magenta, solid, line width=3.0pt, mark=None, mark options={solid, magenta}]
  table[row sep=crcr]{%
0	4.75454083811984\\
10	8.108606279517259\\
20	10.692180470924082\\ 
30	12.9447804785016\\
40	14.8373900216607\\ 
};
\addlegendentry{SIWMMSE \cite{bruno}}

\addplot [color=black, dashdotted, line width=2.0pt, mark=none, mark options={dashdotted, black}]
  table[row sep=crcr]{%
0	3.81977710919928\\
5	5.47600999103445\\
10	6.7780732506182\\
15	7.65574653148277\\
20	8.20674833973187\\
25	8.62478097658254\\
30	9.00648348150387\\
35	9.35926422391261\\
40	9.61003910492672\\
};
\addlegendentry{No RS \cite{amor2020bilinear}}

\end{axis}

\end{tikzpicture}%

%% file: Less_pilot_hs.tex
\definecolor{mycolor1}{rgb}{0.00000,0.44700,0.74100}%
\definecolor{mycolor2}{rgb}{0.85000,0.32500,0.09800}%
\begin{tikzpicture}

\begin{axis}[%
width=0.9\figW,
height=\figH,
at={(0\figW,0\figH)},
scale only axis,
xmin=0,
xmax=40,
xlabel style={font=\color{white!15!black}},
xlabel={$\bf{Transmit\:Power\:}$ $P_{t}$ $\bf{\:in\:dB}$},
ymin=0,
ymax=15,
ylabel style={font=\color{white!15!black}},
ylabel={$\bf{Sum-Rate\:[bpcu]}$},
axis background/.style={fill=white},
xmajorgrids,
ymajorgrids,
legend style={at={(0.95,0.01)}, anchor=south east, legend cell align=left, align=left, draw=white!15!black, row sep=-0.05cm, font=\scriptsize}
]
\addplot [color=mycolor1, solid, line width=3.0pt, mark=diamond, mark options={solid, mycolor1}]
  table[row sep=crcr]{%
0	2.8002016596593\\
5	4.23997737061392\\
10	5.16558301600789\\
15	6.14011327815761\\
20	7.07957861061827\\
25	7.91324863667071\\
30	9.34291791852299\\
35	10.5049274775686\\
40	12.6090662773287\\
};
\addlegendentry{RS BP}

\addplot [color=red, dashdotted, line width=3.0pt, mark=o, mark options={dashdotted, red}]
  table[row sep=crcr]{%
0	2.95647076717583\\
10	5.13477888553116\\
20	7.3948392148803\\
30	9.56073483535811\\
40	12.0310887272465\\
};
\addlegendentry{SINR Inc. Algo \cite{amor2023rate}}

\addplot [color=magenta, solid, line width=3.0pt, mark=None, mark options={solid, magenta}]
  table[row sep=crcr]{%
0	3.73467198066149\\
10	5.96155226188207\\
20	9.01708125155697\\
30	11.7675605186135\\ 
40	14.1927699554785\\ 
};
\addlegendentry{SIWMMSE \cite{bruno}}

\addplot [color=black, dashdotted, line width=2.0pt, mark=none, mark options={dashdotted, black}]
  table[row sep=crcr]{%
0	2.97156113544215\\
5	4.23948662692585\\
10	5.32543582212026\\
15	6.14011327815761\\
20	6.67957861061827\\
25	7.08900642909096\\
30	7.46121072636747\\
35	7.79102532721823\\
40	8.03166142897276\\
};
\addlegendentry{No RS \cite{amor2020bilinear}}

\end{axis}

\end{tikzpicture}%